\documentclass{article}

\usepackage{arxiv}

\usepackage[utf8]{inputenc} 
\usepackage[T1]{fontenc}    
\usepackage{hyperref}       
\usepackage{url}            
\usepackage{booktabs}       
\usepackage{amsfonts}       
\usepackage{nicefrac}       
\usepackage{microtype}      
\usepackage{cleveref}       
\usepackage{lipsum}         
\usepackage{graphicx}
\usepackage{natbib}
\usepackage{doi}
\usepackage{enumitem}

\title{A Lifecycle-Oriented Detection and Defense Framework for Price Manipulation Attacks in DeFi}

\date{}

\newif\ifuniqueAffiliation
\uniqueAffiliationtrue

\ifuniqueAffiliation 
\author{    
    XINGYU XIONG\footnotemark[1]\\
	Hainan University\\
	Haikou, China\\
    \texttt{717349394@qq.com}
    \And
    CHANG LIU\footnotemark[1]\\
	Hainan University\\
	Haikou, China\\
	\texttt{liuchang@hainanu.edu.cn}
    \And
    XIAOQI LI \\
	Hainan University\\
	Haikou, China\\
	\texttt{csxqli@ieee.org}
}

\else
\usepackage{authblk}

\author[1]{%
	\href{https://orcid.org/0000-0000-0000-0000}{\usebox{\orcid}\hspace{1mm}David S.~Hippocampus\thanks{\texttt{hippo@cs.cranberry-lemon.edu}}}%
}
\author[1,2]{%
	\href{https://orcid.org/0000-0000-0000-0000}{\usebox{\orcid}\hspace{1mm}Elias D.~Striatum\thanks{\texttt{stariate@ee.mount-sheikh.edu}}}%
}
\affil[1]{Department of Computer Science, Cranberry-Lemon University, Pittsburgh, PA 15213}
\affil[2]{Department of Electrical Engineering, Mount-Sheikh University, Santa Narimana, Levand}
\fi

\renewcommand{\headeright}{}
\renewcommand{\undertitle}{}
\renewcommand{\shorttitle}{Lifecycle-Oriented Defense for DeFi Price Manipulation}

\hypersetup{
pdftitle={A Lifecycle-Oriented Detection and Defense Framework for Price Manipulation Attacks in DeFi},
pdfsubject={Blockchain Security, Decentralized Finance},
pdfauthor={Chang Liu},
pdfkeywords={Decentralized finance, Oracle, Price manipulation, More},
}

\begin{document}
\maketitle
\footnotetext[1]{These authors contributed equally to this work.}

\begin{abstract}
Aiming at the frequent oracle price manipulation attacks in Decentralized Finance (DeFi), this paper conducts research from three core aspects: attack patterns, vulnerability types, and overall protection schemes.
First, based on the Oracle lifecycle theory, we construct a three-layer attack tree model covering the physical layer, protocol layer, and application layer to analyze the potential impacts on the entire attack chain and identify the data election stage as the optimal intrusion point. Combined with typical attack cases, four main types of vulnerabilities at the contract level are summarized, namely lack of validation on price data flow, oracle call risks, uncontrolled cross-contract calls, and defects in AMM price reading logic.
Second, the fuzzy analytic hierarchy process (Fuzzy-AHP) and value-at-risk (VaR) model are combined to measure the risk magnitude caused by various factors, and a risk matrix consisting of technology, market, governance, and contract is constructed. Subsequently, an automatic detection tool based on the extended Slither framework is developed that uses taint tracking and pattern matching to locate high-risk code.
Finally, to cope with the full-dimensional risk exposure, a multi-layer defense strategy from data source to smart contract to operating system is proposed. At the data source end, a hardware-level trusted execution environment is adopted to ensure the authenticity of the data source. TWAP smoothing and an adaptive circuit breaker are used in the smart contract to mitigate the impact of market fluctuations. The kernel network configuration is optimized at the operating system level to improve system security.
Experiments indicate that this method reduces the price deviation caused by attacks from the original 55.56\% to less than 5\%, which increases the attack cost for hackers. Meanwhile, the tool achieves an accuracy of 94.38\% and a recall rate of 92.31\%, which exceeds the performance of existing open-source software.
\end{abstract}


\section{Introduction}

With the practical deployment of blockchain technology and the rapid growth of decentralized finance (DeFi), smart contracts have become the fundamental operating platform for DeFi protocols \citep{John2023Smart}, while oracles serve as the core components enabling interoperability between on-chain smart contracts and off-chain data. As the assets managed within the DeFi ecosystem continue to grow, so do the security risks it faces, among which price manipulation constitutes one of the most severe attack forms. By tampering with oracle price feeds, controlling the prices of automated market maker (AMM) liquidity pools, or exploiting flash loans to manipulate prices within a single block, attackers can compromise the core functions of DeFi protocols, resulting in substantial financial losses and undermining trust in decentralized finance. Consequently, research on the security threats posed by blockchain price manipulation is both timely and necessary. This paper analyzes the working principles, complete attack process, and resulting harms of price manipulation attacks, and on this basis proposes a price manipulation detection and protection scheme based on the oracle lifecycle, aiming to provide practical insights for building a secure and trustworthy blockchain financial market.

\begin{figure}[htbp]
    \centering
    \includegraphics[width=0.8\textwidth]{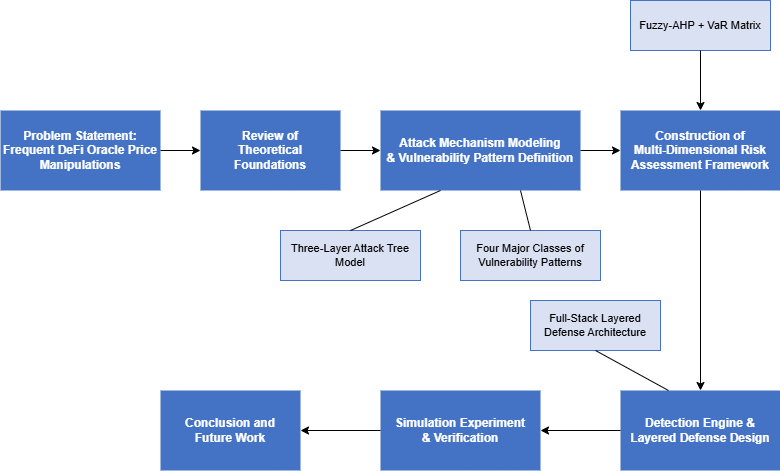}
    \caption{Technical roadmap of this research}
    \label{fig:roadmap}
\end{figure}

Existing research on oracle security and price manipulation can be broadly grouped into three lines of work: lifecycle and attack modeling, empirical analysis of real-world incidents, and detection and defense mechanisms. Regarding lifecycle and attack modeling, Eshghie et al. \citep{eshghie2024oracle} were the first to propose a comprehensive oracle data lifecycle model, dividing the data process into five stages—creation, submission, consensus, election, and elimination—which provides a structured framework for analyzing price manipulation attacks. Based on seven real-world attack cases, they further identified nine distinct attack methods and found that the election stage is where attacks occur most frequently. Okika et al. \citep{Okika2025Smart} adopted a platform-wide perspective, assessing smart contract vulnerabilities and security risks in blockchain-based lending platforms to develop a risk assessment framework that identifies price manipulation as a major risk category.
Regarding empirical evidence, several studies show that price manipulation is rarely an isolated attack vector: Valix's analysis of the Twindex incident \citep{valix2022twindex} showed that attackers combined flash-loan-driven large transactions with price oracle manipulation, while attack databases documenting reentrancy exploits \citep{pcaversaccio2023reentrancy} indicate that reentrancy vulnerabilities are frequently leveraged to alter prices. Public incident datasets \citep{sunweb3sec2023defihacks,wang2026libscan} have further supported this line of empirical research, together suggesting that DeFi attacks increasingly combine multiple vulnerabilities across coordinated steps \citep{li2026systematic}.
Regarding detection and defense, Zhou et al. \citep{Zhou2022SoK} systematized a broad range of DeFi attacks, cataloguing common manipulation techniques such as fake liquidity injection and reviewing existing detection and defense methods, proposing improvements tailored to DeFi's characteristics. Isravel et al. \citep{Isravel2025Reinforcement} proposed a reinforcement-learning-enhanced adaptive oracle framework that aggregates data from multiple sources to improve reliability and reduce the possibility of manipulation at the source. Qian et al. \citep{Qian2023Empirical} conducted an empirical review of smart contract and DeFi security, systematically classifying vulnerability detection techniques and discussing the strengths and limitations of pre-and post-attack detection approaches. From an applied perspective, Arora et al. \citep{Arora2024SecPLF} proposed SecPLF, a secure protocol for loanable funds that identifies and intercepts oracle manipulation attempts via transaction-level safeguards, thereby balancing defense cost against detection accuracy.
Despite this progress, existing studies tend to focus on a single stage of the oracle lifecycle or a single layer of defense, and few works integrate attack-mechanism analysis, vulnerability detection, risk assessment, and multi-layer protection into a unified framework. To address this gap, this paper conducts an in-depth analysis of the operating principles and classification of price oracles together with the attack mechanisms underlying common price manipulation techniques, and proposes a lifecycle-oriented detection and protection scheme. The main contributions of this paper are as follows:

\begin{itemize}
\item \textbf{Analysis of price manipulation attack mechanisms.} We analyze and summarize methods such as oracle data tampering and flash-loan-linked manipulation, identifying their triggering conditions and complete attack processes. Based on this analysis, we extract the main risk points and characterize the relationship between attacks and smart contract code defects \citep{Li2025Penetrating,Wu2024Strengthening,li2025interaction,ding2025comprehensive}.
\item \textbf{Vulnerability patterns and detection methods related to price manipulation.} For representative attack contracts, we perform static analysis to decompose the underlying vulnerabilities and propose four main vulnerability patterns, together with their characteristics, triggering conditions, and risk levels. Taint analysis and pattern matching are then employed to identify these vulnerabilities, yielding an automated detection approach \citep{DiAngelo2023SmartBugs,Kezadri2025Ethereum,li2026psr2,li2026scpatcher}.
\item \textbf{Layered protection measures.} Based on full-process risk analysis and identified contract vulnerabilities, we develop a multi-layer protection scheme spanning the data source layer, the contract layer, and the operating system layer: a hardware root of trust ensures data authenticity at the source layer \citep{Muñoz2023A,Li2023A}. Price smoothing, adaptive weighted aggregation, and circuit breakers are implemented at the contract layer. Kernel-level network tuning compresses the attack window at the operating system layer.
\item \textbf{A price manipulation risk assessment system.} We propose evaluation indicators from the perspectives of technology, market, governance, and contract code vulnerability, and construct a quantitative risk assessment system, which is validated on representative DeFi protocols to identify the main influencing factors and provide a quantitative reference for protective measure design \citep{Kareem2025A,Alfajeer2025stock}.
\item \textbf{Experiments and validation.} We simulate common attacks based on Python implementations to verify the feasibility and timeliness of the proposed defense methods, evaluating them in terms of price control, attack prevention, and system overhead. A test dataset is constructed to compute detection accuracy, recall, and false positive rate, and the results are compared with existing tools to further demonstrate the effectiveness and practicality of the proposed methods.
\end{itemize}
The overall technical roadmap of this research is shown in Figure~\ref{fig:roadmap}.

\section{Related Work}

\subsection{Blockchain and DeFi}
\subsubsection{Blockchain Architecture}
Blockchain is a decentralized distributed ledger organized in a layered architecture, with each layer providing distinct functionalities \citep{long2025fomo3d}. Its basic architecture consists of the data layer, network layer, consensus layer, incentive layer, contract layer, and application layer. The data layer includes block headers, block bodies, hash values, and Merkle trees. The network layer is responsible for data dissemination and verification through a peer-to-peer (P2P) mechanism \citep{zhang2025attacks}. The consensus layer employs consensus algorithms to enable nodes to agree on the validity of block data. The incentive layer maintains the network by rewarding nodes for participating in the bookkeeping process. The contract layer contains smart contracts and scripts, thereby enabling blockchain programmability. The application layer supports a wide range of services, such as digital currencies and decentralized finance. The fundamental technologies underlying blockchain are hash functions, timestamps, and consensus mechanisms. Hash functions ensure data uniqueness and immutability, timestamps preserve the chronological order of transactions, and consensus mechanisms determine which nodes may append new blocks while maintaining network security \citep{Jain2024A,zhou2025blockchain}. The evolution of blockchain technology can be divided into three stages. Blockchain 1.0, represented by Bitcoin, focuses primarily on digital currencies. Blockchain 2.0 is characterized by the introduction of smart contracts, extending applications to the financial and commercial sectors \citep{Zhou2022SoK}. Blockchain 3.0 further expands blockchain applications beyond finance to areas such as social governance and the Internet of Things (IoT).
\subsubsection{Blockchain Consensus Mechanisms}
In blockchain systems, data are linked and stored in a chain structure to ensure immutability. In centralized systems, a central authority exclusively controls system operation and data access. In contrast, all participating nodes in a blockchain network possess equal rights to maintain the ledger. To ensure the authenticity and validity of on-chain data, achieve global data consistency, and prevent malicious nodes from submitting false information, consensus mechanisms have become an indispensable component of blockchain systems. Consensus is, in essence, the process by which multiple participants in a distributed system discuss a given issue and ultimately reach a common agreement. In a centralized environment, the network is governed by a single control point to which all other devices must defer. Blockchain technology, however, adopts a decentralized paradigm. Consequently, a predefined set of rules is required to coordinate all network nodes and ensure that they maintain identical data. Consensus algorithms address this challenge by guaranteeing that the information recorded in the ledgers of distributed nodes remains both consistent and correct \citep{Jain2024A,zhou2025blockchain}. Consensus mechanisms serve two primary functions. First, they validate data to ensure its correctness. Second, they select a node to append new data to the blockchain. Because processing and validating data incur costs, public blockchains typically reward participating nodes with digital tokens to encourage broader participation. Common consensus mechanisms include Proof of Work (PoW), Proof of Stake (PoS), Delegated Proof of Stake (DPoS), and Practical Byzantine Fault Tolerance (PBFT) \citep{Katiyar2023Decentralized, Dagar2025Comparative}.
\subsubsection{Smart Contracts and Decentralized Applications}
Smart contracts are self-executing programs deployed on a blockchain. Once deployed, their internal business logic becomes immutable and can be automatically triggered and executed according to predefined rules. Smart contracts possess three key characteristics. First, determinism: under identical conditions, invoking the same contract always yields the same result, ensuring that on-chain outcomes are verifiable throughout the network. Second, autonomy: after deployment, contracts can execute the prescribed operations automatically whenever the specified conditions are met, thereby enabling decentralized business processes without human intervention. A parallel line of research similarly argues that autonomously executing AI agents require security-by-design rather than reactive patching \citep{li2026defensible}. Third, composability: different smart contracts can interact through standardized interfaces, allowing developers to build diverse on-chain applications upon existing contracts and fostering a highly flexible application ecosystem \citep{Taherdoost2023Smart}. Ethereum was the first public blockchain to support Turing-complete smart contracts. Developers write contract code in the high-level language Solidity and deploy it for execution on-chain through the Ethereum Virtual Machine (EVM) \citep{Tsudzenko2023APPROACHES,gao2025implementation}. Decentralized applications (DApps) leverage the programmability of smart contracts. Although their front-end interfaces resemble those of conventional web applications, their business logic and data are maintained on decentralized blockchains. Among DApps, decentralized finance (DeFi) has emerged as the most important and widely adopted category, encompassing decentralized exchanges, lending platforms, derivatives, asset management, and a variety of other financial products and services that constitute the primary real-world applications of blockchain technology today \citep{Kareem2025A, John2023Smart}.
\subsubsection{Automated Market Makers and Liquidity Pools}
The Automated Market Maker (AMM) is the key innovation underlying decentralized exchanges (DEXs), replacing the traditional order-book mechanism with smart contracts. Unlike centralized exchanges, which rely on matching buy and sell orders, AMMs employ algorithmic pricing formulas and liquidity pools to facilitate asset trading without requiring direct counterparties \citep{bagnulo2025poolingliquiditypoolsamms}.AMM liquidity pools operate according to predefined mathematical rules, the most common of which is the constant-product formula, $x \times y = k$, where x and y denote the quantities of the two assets in the pool and k is a constant that remains invariant \citep{philippe2023amm, bronnimann2024amm}. When users execute trades, they interact directly with the liquidity pool rather than being matched with other traders \citep{chu2026impermanent}.

\subsubsection{Flash Loans}
Flash loans eliminate the traditional requirement for collateralized assets, thereby substantially lowering the barriers to arbitrage, asset reallocation, and self-liquidation \citep{warodom2022flashloan}. Fundamentally, a flash loan consists of a sequence of programmable operations whose successful execution relies on the final-state validation mechanism of virtual machines such as the Ethereum Virtual Machine (EVM) \citep{li2026atomgraph}. Consequently, flash loans have also been exploited by attackers and have been involved in numerous DeFi security incidents, making them a central topic in blockchain security research \citep{Gao2025Flash}. In protocols such as Aave, any developer may invoke a smart contract to borrow assets, provided that the principal and the associated fee are repaid within the same transaction. Otherwise, the entire transaction is reverted automatically, and the funds are returned to the liquidity pool.

\subsection{Oracles and Price Data Mechanisms}
\subsubsection{Definition and Functions of Oracles}
An oracle serves as the bridge between the closed blockchain environment and the external world by providing off-chain data to on-chain applications. Because blockchain networks operate as self-contained systems and cannot natively access real-world information, oracles have become indispensable infrastructure for DeFi protocols to obtain external data and execute on-chain business logic \citep{you2024persona, Caldarelli2021The}.\
In a complete data pipeline, an oracle performs the entire process from data acquisition to continuous updates. It first retrieves raw market data from multiple sources, such as exchange APIs and price aggregators. The heterogeneous data are then cleaned and aggregated to compute the final price feed. The verified results are subsequently submitted to on-chain smart contracts for use by DeFi protocols. Moreover, the oracle periodically updates the on-chain data according to predefined rules, ensuring that the reported prices accurately reflect current market conditions \citep{caldarelli2022overview}.\
The security and accuracy of oracle data directly determine the safety of the assets managed by DeFi protocols that rely on such price feeds and thus constitute a fundamental guarantee for the stable operation of the DeFi ecosystem. If an oracle is maliciously manipulated, critical on-chain operations—including lending, trading, and liquidation—may be executed based on erroneous price information, potentially resulting in substantial financial losses for both protocols and users \citep{pasdar2023connect}. Consequently, oracle security has become a central research topic in the field of DeFi security.

\subsubsection{Mainstream Oracle Solutions}
\begin{enumerate}[label=(\arabic*), leftmargin=16pt]
\item \textbf{Chainlink}

Within blockchain infrastructure, Layer-1 networks, cross-chain bridges, and consensus mechanisms have long been focal points of both academic research and market attention. However, there exists another category of infrastructure that serves as a fundamental technological underpinning and constitutes the cornerstone of the healthy development of the blockchain ecosystem. The value of such infrastructure is often deeply embedded in ecosystem operations and therefore easily overlooked. Chainlink is a representative example of this category. As the leading decentralized oracle network in today's DeFi ecosystem, Chainlink provides a reliable and secure means of connecting external data sources to blockchain smart contracts through a consensus-based, multi-node verification design, thereby supporting a wide range of blockchain applications \citep{yao2025pscbo, kaleem2021demystifyingpythiasurveychainlink}. By 2025, Chainlink had been deployed on more than fifteen public blockchains, providing security assurances for the DeFi market with a total value exceeding USD 200 billion. Chainlink employs a decentralized network of nodes to bring off-chain data on-chain. Specifically, geographically distributed nodes retrieve the latest information from the real world, and after the authenticity and validity of the data are verified through a consensus mechanism, the information is packaged into standardized data feeds and delivered on-chain for invocation by smart contracts \citep{Gupta2023Proxy}. In this way, Chainlink overcomes the inherent limitation that smart contracts cannot actively access off-chain data. By bridging Ethereum and the real world, Chainlink enables the operation of DeFi protocols, blockchain games, and a wide variety of smart-contract-based applications that would otherwise be impractical without reliable off-chain data support.

\item \textbf{Pyth Network}

Pyth Network is a next-generation blockchain oracle project built on the Solana ecosystem with its own dedicated chain architecture \citep{wu2025exploring}. Its core mechanism is a pull-based price update model, in which price feeds are updated on demand in response to client requests. Compared with traditional oracle solutions, Pyth exhibits significant advantages in three key aspects. First, it offers superior update speed, supporting high-frequency price updates with latencies as low as 300--400 ms. Second, it enjoys extensive coverage, integrating real-time price feeds from more than 90 data providers and serving over 50 blockchains, 144 service networks, and 162 partner protocols, making it second only to Chainlink in overall scale. Third, its price feeds are characterized by high data fidelity, thereby enhancing pricing accuracy. In terms of tokenomics, 22\% and 52\% of the token supply are allocated as incentives for data providers and contributors to ecosystem development, respectively, effectively encouraging participation from both groups. Consequently, Pyth has experienced rapid growth. It secures approximately USD 5.5 billion in assets across 162 protocols deployed on more than 50 blockchains, ranking behind only Chainlink among oracle networks. In February 2024, transactions involving the Pyth oracle accounted for, on average, 20\% of Solana's total transaction volume, while Pyth data providers on Solana received approximately USD 225,000 in fees during the same month. By May 6, 2024, the cumulative trading volume associated with Pyth Data had reached USD 2 billion.

\item \textbf{Band Protocol}

Band Protocol is also a decentralized oracle project dedicated to addressing the single point of failure and trust issues inherent in traditional centralized oracles. In conventional oracle models, data sources typically rely on a single provider or a small number of providers, which not only limits the diversity and reliability of the data but also increases the risk of data tampering and manipulation \citep{Aspembitova2022Oracles}. However, compared to Chainlink, its ecosystem scale remains relatively small, and the richness of its ecosystem still needs to be improved.
\end{enumerate}
\subsubsection{Price Feed Mechanisms and Data Aggregation}
Price feed mechanisms and data aggregation constitute the two core components of oracle solutions \citep{you2024persona, Pasdar2022Connect}. Price feed mechanisms generally adopt either a push-based or a pull-based approach. The push-based model, exemplified by Chainlink, relies on nodes that continuously monitor off-chain data sources and proactively submit aggregated data to on-chain contracts whenever the price deviation exceeds a predefined threshold or a heartbeat interval elapses \citep{Gangwal2022Analyzing}. Although this approach is developer-friendly, it entails higher costs and longer update latencies. In contrast, the pull-based model, represented by Pyth, requires nodes only to sign and verify off-chain data. The verified data are then fetched on-chain by users when needed. This approach offers lower latency and pay-as-you-go pricing, but users must bear the cost of data retrieval themselves \citep{Gansäuer2025Price}.\
Data aggregation mechanisms enhance data accuracy and resistance to manipulation by combining information provided by multiple independent nodes or data sources \citep{Isravel2025Reinforcement}. Common aggregation methods include using the median to filter out outliers, computing weighted averages based on node staking weights, and assigning weights according to reputation scores derived from historical performance \citep{caldarelli2021blockchain}. The design of the aggregation layer directly affects the robustness of the final price feed against manipulation and determines the overall quality of the oracle data.

\subsection{Oracle Lifecycle Theory}
The oracle lifecycle theory divides the entire process of an oracle, from data acquisition to final output, into five stages, each of which is associated with specific security risks.
\subsubsection{Data Creation}
The data creation stage encompasses raw data acquisition and preprocessing. If the data sources used by an oracle are manipulated or contain erroneous information, the entire oracle pipeline will be compromised, since the oracle's output depends directly on the authenticity of its input data \citep{Deng2024Safeguarding}. In addition, a low data collection frequency may cause delays in reflecting market price changes, thereby creating a time window that attackers can exploit for arbitrage or price manipulation \citep{valix2022twindex}. Improperly designed outlier filtering algorithms may also discard valid data, causing the final price feed to deviate from actual market conditions \citep{Aspembitova2022Oracles}.
\subsubsection{Data Submission}
During the data submission stage, nodes submit their observations to on-chain contracts. Malicious nodes may intentionally provide false data in an attempt to manipulate the final price feed for profit. Moreover, network latency may cause some nodes to submit stale data, resulting in poor timeliness. If the submission times of different nodes within a single aggregation round differ significantly, attackers may launch Sybil attacks by controlling multiple seemingly independent nodes, thereby increasing the weight of malicious submissions in the aggregation process \citep{you2024persona}.
\subsubsection{Data Consensus}
In the data consensus stage, the data submitted by multiple nodes are aggregated to generate the final price feed. Aggregation algorithms based on simple averaging or medians may be exploited by attackers through the submission of carefully crafted values. If an attacker gains control of a majority of the participating nodes, the final price feed can be directly manipulated to deviate from the true market price \citep{wu2024defiranger}. Furthermore, discrepancies among the timestamps of data submitted by different nodes may affect the correctness of the aggregation result, depending on how such temporal differences are handled \citep{Liu2023A}.
\subsubsection{Data Election}
The data election stage determines which nodes are eligible to participate in data submission. Eshghie et al. argue that this stage is particularly susceptible to attacks. If the node selection process relies on predictable randomness, attackers may anticipate the election outcome and strategically deploy malicious nodes in advance \citep{Gorman2025VRaaS}. Moreover, when the amount of tokens staked by participating nodes is insufficient, the cost of misconduct becomes too low to effectively deter malicious behavior \citep{eshghie2024oracle}. In voting-based schemes such as Delegated Proof of Stake (DPoS), attackers may further manipulate the voting process to influence the election outcome, thereby enabling malicious nodes to gain the right to participate in data submission \citep{yao2025pscbo}.
\subsubsection{Data Elimination}
The data elimination stage concerns the management of node departures and identity changes. If a malicious node that has exited the network is still allowed to participate in data submission for a certain period, the effectiveness of the elimination mechanism is directly undermined \citep{Ningthoujam2025Blockchain}. Furthermore, removed nodes may rejoin the network under new identities, thereby evading penalties associated with their previous misconduct. Some oracle designs also permit modifications to historical data. Such a feature not only complicates auditing procedures but may also conceal traces of attacks, ultimately reducing the overall trustworthiness of the system.

\subsection{Smart Contract Security}
Smart contract security analysis provides the key technical foundation for identifying price-manipulation vulnerabilities in DeFi protocols and serves as an important basis for the vulnerability taxonomy and detection methods proposed in this paper \citep{li2026uscsa}. Since most oracle and automated market maker (AMM) contracts in DeFi are implemented in Solidity, they are first compiled into abstract syntax trees (ASTs) and subsequently translated into EVM stack-based bytecode for on-chain deployment. In the absence of source code, bytecode decompilation can be employed to recover the contract execution flow, thereby enabling contract analysis and vulnerability localization.\
Formal verification, symbolic execution, taint analysis, and intermediate-representation-based detection techniques are all performed without executing the target program. Accordingly, these approaches are collectively classified as static analysis methods, which primarily include control-flow graph construction, data-flow analysis, taint propagation, and pattern matching \citep{Kezadri2025Ethereum,li2026ckg}. Representative tools include the open-source static analysis framework Slither, implemented in Python. Mythril, which leverages symbolic execution. and Oyente, which specializes in the analysis of EVM bytecode, alongside execution frameworks such as SmartBugs that integrate multiple weakness-detection tools within a unified pipeline \citep{DiAngelo2023SmartBugs}.

\subsection{Operating-System-Level Isolated Computing and Network Characteristics}
\subsubsection{Core Definition}
Operating-system-level isolated computing, also known as operating-system-level virtualization, is a technology that leverages the native functionality of the host operating system kernel to create multiple mutually independent isolated execution environments within a single instance \citep{Ketha2025Analysis}. Compared with hardware-level full virtualization technology, its main distinction lies in the absence of a virtualized hardware layer and the elimination of the need to install an additional guest operating system. Instead, it only requires sharing the host kernel, while allocating an independent user space to each isolated environment. This achieves fine-grained control over computing, storage, networking, and other resources, and constitutes an important underlying technology for cloud-native and containerized applications today \citep{Theodoropoulos2023Security}.
\subsubsection{Core Underlying Technologies}
The underlying core capabilities of this technology are jointly supported by three native kernel modules, which work in concert to achieve comprehensive protection of the isolated environment. As the foundation of isolation, namespace is a fundamental feature of the Linux kernel that assigns an independent system view to each isolated unit, achieving complete logical isolation between different isolated units in terms of processes, file systems, inter-process communication, networking, permissions, and other aspects. Cgroup is the key component responsible for resource management. It enables precise quota allocation, priority management, and usage monitoring of various system resources within an isolated unit—such as CPU, memory, disk I/O, and network bandwidth—and can promptly interrupt services in the event of anomalies, thereby ensuring resource fairness among multiple concurrently running isolated units as well as the robustness of the overall host system. In addition, a number of security-hardening measures, such as the minimal-privilege decomposition of Capabilities, whitelist-based filtering of system calls via Seccomp, and mandatory access control, further reduce the system's attack surface and mitigate the security risks arising from kernel sharing, thereby strengthening the security of the isolated environment \citep{Jarkas2025A}.
\subsubsection{Core Computational Characteristics}
The main computational characteristics of this technology can be summarized in four aspects. The first is lightweight deployment and high-concurrency capability. Since no separate guest operating system is required, the additional overhead is eliminated, enabling near-instantaneous startup with minimal resource consumption. As a result, the number of instances that can be deployed on a single host far exceeds that achievable with traditional hardware virtualization approaches. The second is efficient utilization of host resources and elastic scalability. This technology can substantially improve the utilization rate of system resources on the host and achieve second-level elastic scaling, meeting the scheduling requirements of the cloud-native era. The third is environmental consistency and cross-platform portability. Applications together with all the dependencies required for their execution can be packaged as a whole and run normally across different hosts, making this technology an important tool for DevOps as well as continuous integration and continuous delivery \citep{Sultan2023CONTAINER-BASED}. The last is good compatibility and security. It naturally supports mainstream hardware platforms such as x86 and ARM, and on this basis can further be combined with corresponding security measures to satisfy requirements for both operational efficiency and security.
\subsubsection{Core Network Characteristics}
The networking capability of this technology is implemented based on the Network Namespace in the Linux kernel, with each isolated zone possessing an independent and complete network protocol stack, thereby achieving effective network isolation \citep{Qi2021Assessing}. Its main characteristics are as follows. First, strong network isolation: it prevents port conflicts among different isolated zones and blocks unauthorized data flows within an isolated zone, thereby satisfying network isolation requirements among multiple tenants. Second, multiple network interconnection modes: five different network connection modes—Bridge, Host, Overlay, Underlay, and None—are available, which can essentially address networking requirements under almost all network environments. Third, fine-grained traffic control capability: bandwidth can be limited, traffic can be shaped, and access control rules can be configured. In addition, all traffic can be monitored and traced \citep{Budigiri2021Network}. Fourth, high efficiency and good scalability: there is no additional virtualization forwarding overhead, and the network forwarding performance approaches that of the host itself. The technology supports pluggable extensions through standard CNI plugins, allowing the use of various high-performance networking solutions. Fifth, enhanced security: encrypted transmission is supported for all traffic, providing comprehensive protection for the security of all traffic, and the technology can also be integrated with intrusion detection systems to defend against network-layer intrusions.

\section{Method}

\subsection{Attack Mechanisms and Vulnerability Patterns}
\subsubsection{Formal Logic of Price Manipulation Attacks}
The core of a price manipulation attack lies in disrupting the asset equilibrium of an AMM liquidity pool, which is governed by the constant-product formula $x \times y = k$. By injecting a large amount of asset $x$ into the pool, an attacker causes the price of asset $y$ to deviate substantially from the normal market price within a short period, creating room for subsequent profit-taking. A flash-loan-based price manipulation attack generally proceeds through four stages: (i) acquiring the initial attack capital, (ii) distorting the price equilibrium of the AMM liquidity pool, (iii) inducing the oracle to read the anomalous price and feed it on-chain, and (iv) extracting excess collateral value based on the anomalous price to complete the arbitrage loop.

\subsubsection{Systematic Design of the Attack Structure}
\begin{figure}[htbp]
    \centering
    \includegraphics[width=0.8\textwidth]{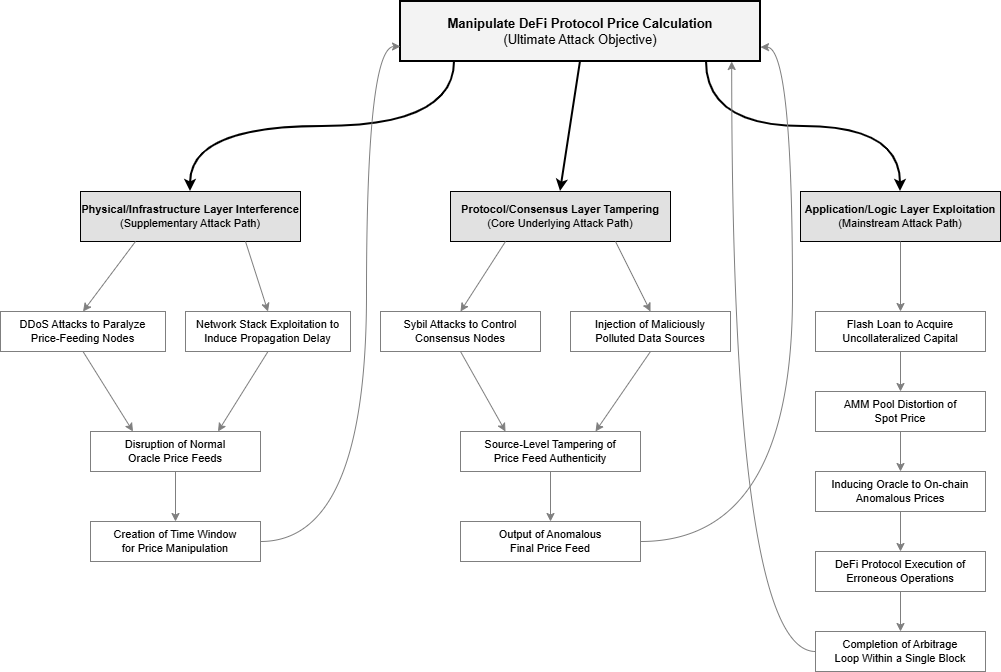}
    \caption{Three-layer attack tree model.}
    \label{fig:attack_tree}
\end{figure}

Taking DeFi protocol price manipulation as the root node, this paper constructs a three-layer attack tree across the entire technology stack, based on the hierarchical structure of the blockchain technology stack and the propagation paths of attacks (as shown in Figure~\ref{fig:attack_tree}). This attack tree divides the attack path into three layers—physical and infrastructure-layer interference, protocol and consensus-layer tampering, and application and logic-layer exploitation—corresponding respectively to an auxiliary attack path, an underlying core attack path, and a common industry path. It illustrates the process by which price manipulation attacks occur, and further incorporates multi-path joint attack modes, covering essentially all of the major price manipulation techniques currently present in the DeFi domain.\\

Physical and infrastructure-layer attacks aim to disrupt the operating environment of nodes and serve as an effective supplementary means of price manipulation, functioning to reserve a time window and create a favorable environment for subsequent attacks. An attacker can launch a DDoS attack against the core price-feeding nodes of an oracle to render them inoperative, or introduce transmission delays at the underlying network stack of the operating system. Both methods prevent the oracle node from feeding prices on-chain in a timely and accurate manner, thereby creating a discrepancy between the on-chain price and the actual market price and, in turn, an opportunity for price manipulation.\\

Protocol and consensus-layer attacks aim to tamper with the source of price-feed data and represent the most fundamental pathway for price manipulation attacks, intervening in oracle node selection, consensus, and data generation during the price-feeding process so as to fundamentally affect the authenticity of the feed. An attacker may launch a Sybil attack during node election, forging a large number of fake nodes to enter the network and control the majority of consensus nodes in order to alter the final price-feed result. Alternatively, the attacker may exploit the lack of multi-party cross-validation design in some oracles by implanting malicious data sources and falsified price information into part of the oracle system, thereby interfering with the basis of price-feed computation at the source and causing the on-chain price to deviate severely from the genuine market price.\\

Application and logic-layer attacks aim to exploit on-chain business rules and represent the most common form of price manipulation currently observed in DeFi. They also constitute the key step for completing the entire attack loop and realizing profit. Exploiting the atomicity of flash loans, an attacker can complete the entire attack within a single block: first borrowing a large amount of uncollateralized funds via a flash loan, then injecting all of it at once into the AMM liquidity pool of the target trading pair, causing an imbalance in the asset ratio within the liquidity pool and thereby affecting the market price. The oracle then reads this manipulated price and feeds it on-chain. Since many DeFi projects whose functions—such as liquidation or lending—depend on oracle-derived values are affected by this price, once the attacker has secured a profit, the principal and interest of the flash loan can be repaid within the same transaction, avoiding losses arising from price changes across blocks, and thereby achieving cost-free or even low-cost price manipulation and arbitrage.\\

The three attack-layer paths can achieve the objective of price manipulation either independently or in combination to form a multi-layered attack approach, thereby increasing both the probability of success and the degree of concealment. Together, they comprehensively cover the major technical means of price manipulation currently present in the DeFi ecosystem, while also pointing the way toward the construction of a future all-around, multi-layered defense system.

\subsubsection{Attack Vectors Across Stages}
From the perspective of attack-path correlation, this paper considers the oracle's data election stage to be the optimal injection point for an attacker to carry out a price manipulation attack. Node election and data consensus within the oracle network fundamentally rely on the underlying Verifiable Random Function (VRF). If the underlying node's genuine random number generation process lacks the protection of a hardware root of trust, an attacker can exploit algorithmic flaws to predict the outcome of the random number, thereby locking in and controlling the price-feeding nodes for the next round beforehand, so as to manipulate the resulting oracle price feed and lay the groundwork for subsequent price manipulation.

\subsubsection{Smart Contract Vulnerability Patterns}
Building on an analysis of the vulnerability origins of typical DeFi price manipulation attack incidents, and based on the core risk targets identified by the attack tree model, this section classifies and defines four core vulnerability patterns in contract code that may lead to price manipulation, describing the code characteristics, trigger conditions, and degree of harm associated with each pattern, so as to provide direction for subsequent research. Risk levels can be classified as high risk, medium risk, and low risk.\\
The first pattern is the category of missing price data flow validation. This category of pattern is the most common type of vulnerability in price manipulation attacks. Its core characteristic is the absence of necessary validity and reasonableness checks throughout the entire process of oracle price data, from acquisition to business application. An attacker can artificially tamper with the spot price so that the contract directly adopts the anomalous data to realize arbitrage. This is the core high-risk vulnerability type that triggers such attacks.\\
\begin{enumerate}[label=(\arabic*), leftmargin=16pt]
    \item Direct single-block price reading without smoothing. The core code characteristic of this pattern is that the contract directly reads the instantaneous spot price of the AMM pool for the current block, without applying TWAP (time-weighted average price) smoothing, and without any multi-block price dilution mechanism. Under such conditions, an attacker can use a flash loan to complete a large-value transaction within a single block to distort the spot price, and the contract directly reads this anomalous price for core business functions such as liquidation and collateral valuation. This is classified as high risk.

    \item Absence of price deviation threshold validation. The core code characteristic of this pattern is that the contract imposes no threshold limit whatsoever on the deviation between an updated price feed and the historical benchmark price, accepting the update regardless of how large the price change is. When a malicious party manipulates the price to cause a substantial deviation, the contract also lacks any circuit-breaker mechanism to prevent the anomalous price from being used in subsequent operations. The risk level is relatively high.

    \item Defective outlier filtering logic. The core code characteristic is that the contract computes the price feed submitted by multiple nodes using only a simple average, without removing outliers through methods such as the median or a trimmed mean, so that extreme anomalous data can directly affect the final price-feed result. The trigger condition occurs when an attacker controls some of the price-feeding nodes and submits extreme anomalous prices, thereby pulling the final aggregated price down or up through the simple averaging algorithm to achieve price-feed manipulation. This is classified as medium risk.

    \item Absence of multi-source data cross-validation. The core code characteristic is that the contract uses the price information from only a single oracle or a single AMM pool as its sole price-feed source, without any other data source for cross-validation. Under such circumstances, if an attacker manipulates the price of that single data source, the contract has no alternative validation channel and will directly adopt the manipulated price data. This is classified as a high-risk situation.
\end{enumerate}

The second pattern is the oracle invocation risk category. This category of vulnerability patterns focuses on flaws in oracle invocation logic and access control. Attackers may achieve price manipulation by tampering with oracle call targets or illicitly triggering price-feed updates.\\
\begin{enumerate}[label=(\arabic*), leftmargin=16pt]
    \item Lack of access control for price-feed update functions. The key code characteristic is that the oracle price update function is declared with \verb+external+ visibility and imposes no restrictions on the caller, allowing any address to trigger a price update. The vulnerability is exploitable when an attacker actively invokes the update function within the same block in which the AMM price has been manipulated, causing the contract to immediately read the manipulated price. This enables the attacker to precisely control the timing of the oracle update and constitutes a high-risk vulnerability.

    \item Malicious modification of the oracle address. The key code characteristic is that the state variable storing the oracle contract address is not immutable, while the function for changing this address lacks stringent access controls such as multi-signature authorization or time locks. The vulnerability is triggered when an attacker, through private-key compromise or privilege escalation, changes the oracle address to that of a malicious contract under the attacker's control. The target contract then directly consumes falsified prices, making this a high-risk vulnerability.

    \item Absence of timestamp validation for price-feed data. The key code characteristic is that the contract fails to verify the freshness of the timestamp associated with oracle data and thus accepts stale prices that fall outside the valid time window. During periods of significant market volatility, attackers may exploit the use of outdated prices to conduct arbitrage attacks. This is classified as a medium-risk vulnerability.
\end{enumerate}

The third pattern is the uncontrolled cross-contract invocation category. The core issue of this category is that, during cross-contract retrieval of price data, the contract fails to validate the legitimacy of the data source or the correctness of returned values, allowing malicious contracts to intercept or tamper with price information.\\
\begin{enumerate}[label=(\arabic*), leftmargin=16pt]
    \item Lack of validation of the calling source. The key code characteristic is that the contract function responsible for receiving price data does not verify the caller's address, allowing any contract to submit price information. An attacker can therefore deploy a malicious contract and submit falsified prices, which are accepted without verification, resulting in price manipulation. This is a high-risk vulnerability.

    \item Failure to validate return values from external contracts. The key code characteristic is that the contract obtains prices from an external oracle via \verb+staticcall+ without checking whether the call succeeds or whether the returned data are valid. If the oracle invocation fails, a default value of zero or other erroneous data may be returned and used directly by the contract, leading to incorrect pricing that can be exploited for arbitrage. This is a medium-risk vulnerability.

    \item Price-data hijacking by malicious contracts. The key code characteristic is that the contract employs a proxy pattern to query an external contract for price information without verifying the logic of the target implementation. Consequently, a malicious implementation behind the proxy may intercept and manipulate price data. The vulnerability is triggered when an attacker upgrades the implementation of the oracle proxy contract and alters the returned prices, causing the contract to consume tampered price data. This is a high-risk vulnerability.
\end{enumerate}

The fourth pattern is the AMM price retrieval logic defect category. The core issue of this category of vulnerability patterns is that the contract logic for retrieving prices from AMM liquidity pools is flawed, as it fails to account for market factors such as liquidity depth and slippage. Consequently, the price data can be readily manipulated through large trades.\\
\begin{enumerate}[label=(\arabic*), leftmargin=16pt]
    \item Price retrieval without liquidity-depth validation. The key code characteristic is that the contract computes prices directly from the reserve balances of an AMM pool without verifying the pool's liquidity depth, i.e., its total value locked (TVL). Regardless of the amount of locked liquidity, the computed price is accepted unconditionally. The vulnerability is triggered when an attacker targets a low-liquidity AMM pool, where a small capital injection can substantially distort the price. Since the contract performs no liquidity check, it directly adopts the manipulated price, enabling low-cost price manipulation. This is a high-risk vulnerability.

    \item Use of instantaneous prices without accounting for slippage. The key characteristic is that the contract ignores the slippage caused by large trades and directly uses the post-trade marginal price as the global price feed, without reference to the fair value within the pool. The vulnerability is triggered when an attacker executes a large transaction that induces substantial slippage and distorts the marginal price. The contract then adopts this abnormal price as the global price feed. This is a high-risk vulnerability.

    \item Using the price of a single pool as the global price feed. The key characteristic is that the contract directly uses the price data from a single AMM trading pool as the platform-wide price feed, rather than aggregating prices from multiple major pools. Consequently, an attacker needs only manipulate the price of a single pool to control the platform's global price feed, thereby reducing the cost of the attack. This is a high-risk vulnerability.
\end{enumerate}

\subsubsection{Contract Vulnerabilities in Attack}
The contract vulnerabilities exploited in the Warp Finance attack illustrate this risk. Warp Finance was a decentralized lending platform on Ethereum that was compromised by a flash-loan-driven price manipulation attack in December 2020. The attacker ultimately gained approximately USD 7.8 million, making this incident one of the earliest and most notable examples of an AMM oracle price manipulation attack in DeFi. From the perspective of the contract code, the success of the attack can be attributed to several high-risk security flaws in Warp Finance's pricing contracts. First, the protocol directly consumed the spot price of the current block from a Uniswap pool without any smoothing mechanism such as a TWAP, enabling the attacker to manipulate the price within a single block using a flash loan. Second, the contract neither verified the liquidity depth of the pool nor cross-validated prices using multiple data sources. By relying exclusively on a single Uniswap pool and ignoring its liquidity level, the protocol substantially reduced the attack cost. This incident demonstrates the severity of vulnerabilities related to insufficient price-data validation and flaws in AMM price retrieval logic. These design deficiencies allowed the attacker to perform low-cost price manipulation without requiring upfront capital.\\
A similar pattern of contract vulnerabilities can be observed in the Twindex attack. Twindex was a decentralized synthetic asset protocol deployed on Binance Smart Chain. In February 2022, it suffered a price manipulation attack in which the attacker exploited flash loans to manipulate market prices and obtain a profit of approximately USD 3.8 million. The attack resulted from the combination of multiple vulnerabilities. Examination of the contract code reveals two principal weaknesses in the Twindex oracle. First, the price-feed update function was publicly callable and imposed no access restrictions, allowing the attacker to trigger a price update within the same block in which the market price had been manipulated. Second, the contract lacked a price-deviation check. No upper bound was imposed on acceptable price deviations, so prices exceeding the normal range by more than 50 percent could still be treated as valid. This case illustrates that weaknesses related to oracle invocation and inadequate validation of price data streams can interact to facilitate successful attacks. Notably, the vulnerability categories proposed in this paper comprehensively cover all of the key flaws exploited in this real-world incident.

\subsection{Price Manipulation Risk Assessment Framework}

\subsubsection{Construction of the Evaluation Indicator System}
This study employs the Fuzzy Analytic Hierarchy Process (Fuzzy-AHP) to address the integration of qualitative and quantitative indicators, as well as the uncertainty associated with their relative importance in DeFi risk assessment. On this basis, a comprehensive price manipulation risk assessment framework is established, incorporating technical, market, governance, and smart-contract vulnerability factors.\\
The technical dimension primarily evaluates the robustness of the underlying oracle infrastructure and its ability to withstand attacks. Three indicators are considered. The first is the price-feed heartbeat rate, which is directly related to the attack window. A lower heartbeat frequency provides attackers with more time to carry out malicious operations. The second is the geographical dispersion of oracle nodes. Greater concentration of node locations increases the risk of regional failures and collusion among nodes. The third is the degree of operating-system-level isolation employed by the nodes. Nodes running within trusted execution environments (TEEs) or other isolated environments exhibit stronger tamper resistance and greater resilience to attacks.\\
The market dimension focuses on the ability of AMM liquidity pools to resist price shocks, which fundamentally determines the cost of a price manipulation attack. Three indicators are considered. The first is liquidity depth: the lower the total value locked (TVL), the less capital is required to move the asset price. The second is the asset slippage index. Higher slippage implies that large trades exert a greater impact on the spot market, making asset prices easier to manipulate. The third is the volatility of AMM reserves. Greater reserve volatility indicates less stable liquidity within the AMM and creates more opportunities for exploitation by attackers.\\
The governance dimension assesses the security and decentralization of protocol governance, which constitute the root causes of price manipulation risk. Two key indicators are considered. The first is the Gini coefficient of DAO governance token ownership. Greater concentration of token holdings makes protocol governance more susceptible to control by a small number of participants. The second is the presence of a multi-signature mechanism for administrator privileges. Multi-signature authorization can substantially reduce the risk of malicious parameter modifications resulting from the compromise of a single private key.\\
The smart contract vulnerability dimension constitutes a central component of the proposed evaluation framework. It assesses the resilience of DeFi smart contracts against price manipulation, which represents the most direct cause of such attacks. Four aspects are considered: the comprehensiveness of price-data validation, the strictness of oracle invocation access control, the adequacy of security checks in cross-contract interactions, and the soundness of the AMM pricing mechanism. Quantitative scores are assigned accordingly. A higher degree of validation and protection yields a higher score and, consequently, a lower risk of price manipulation.

\subsubsection{Risk Level Matrix Evaluation}
This study employs the Value at Risk (VaR) model, which is widely used in finance, to quantify the economic losses caused by price manipulation attacks. The weights of the indicators are then determined using the Fuzzy-AHP method, and a weighted aggregation is performed to obtain the overall risk score of a DeFi protocol. Finally, a standardized risk-level matrix is constructed to enable horizontal comparison and evaluation across multiple protocols.\\
The results indicate that protocols with highly concentrated governance structures exhibit greater vulnerability during large-scale market liquidations, demonstrating that the governance dimension is the most fundamental determinant of price manipulation risk. Excessive concentration of governance power may prevent a protocol from responding promptly to extreme market conditions and adjusting its risk parameters. It may even facilitate malicious behavior by insiders or collusion with external attackers to carry out price manipulation, thereby further increasing the protocol's risk exposure.\\
The proposed framework can be directly applied to ex ante risk screening of DeFi protocols, dynamic attack-risk assessment during operation, and the provision of quantitative guidance for investors seeking lower-risk protocols and for regulators formulating risk management standards for the DeFi industry.

\subsection{Design of Defense Framework}
Based on the foregoing analysis of the mechanisms of price manipulation attacks, the classification criteria for contract vulnerabilities, and the corresponding risk assessment methodology, this section proposes an automated approach for detecting smart contract vulnerabilities related to price manipulation, thereby enabling the early identification and targeted inspection of potential risks. Furthermore, for all identified risks and vulnerabilities throughout the attack chain, a comprehensive, multi-layered defense framework is developed at the data source, contract, and operating system levels. By leveraging trusted execution environments and kernel-level hardening techniques, the framework aims to provide ex ante warning, real-time mitigation, and ex post traceability, thereby fundamentally reducing the exposure of DeFi oracles to price manipulation attacks.

\subsubsection{Identifying Price-Manipulation-Related Vulnerability}
Existing mainstream smart contract vulnerability detection tools suffer from limited rule coverage, high false-positive rates, and insufficient specificity when applied to DeFi price-manipulation vulnerabilities. Accordingly, based on the four categories of vulnerabilities proposed earlier, a dedicated identification method is designed to support the subsequent defense framework and to facilitate early-stage risk scanning during DeFi protocol development.\\
With regard to the design objectives and overall workflow, the proposed method pursues four objectives. First, it seeks to comprehensively cover all categories of price-manipulation-related vulnerabilities identified in this study and to detect high-risk flaws accurately and efficiently. Second, it employs taint analysis to monitor all price-related data flows throughout the contract execution process, thereby addressing the inability of conventional approaches to effectively identify missing validation of price data streams. Third, it aims to fully automate vulnerability detection and risk assessment, ultimately generating a standardized report together with mitigation recommendations. Fourth, it is designed to be implemented using open-source tools, making it suitable for practical security auditing during DeFi protocol development.\\
The proposed vulnerability pattern identification method consists of contract preprocessing, control-flow and data-flow construction, taint labeling of price-related data flows, vulnerability pattern matching, and risk-level reporting, forming a complete automated detection pipeline. The process begins by obtaining the Solidity source code or EVM bytecode of the target contract. Lexical and syntactic analyses are then performed to construct the abstract syntax tree, while bytecode is decompiled to recover the contract's principal functionality and basic information. Based on the preprocessing results, a control-flow graph is generated to identify basic blocks and branch transitions. Simultaneously, data-flow analysis is conducted to determine the declaration, propagation, and use of state and local variables, thereby facilitating subsequent taint tracking. Price data originating from AMM pools and external oracle contracts are then designated as taint sources. A forward taint-tracking algorithm is employed to trace the propagation of such data throughout the contract and to identify tainted paths lacking adequate validation. Finally, the four categories of vulnerability patterns proposed earlier are used as the rule set for pattern matching. The preprocessing, data-flow analysis, and taint-tracking results are compared against these rules to identify price-manipulation vulnerabilities. The number, severity, and exploitability of the detected issues are subsequently combined with the risk assessment methodology introduced earlier to determine the contract's overall level of price manipulation risk and to generate a standardized security report. Building on this workflow, the core detection rules are designed as follows, covering four categories of price-manipulation-related vulnerabilities:
\begin{enumerate}[label=(\arabic*), leftmargin=16pt]
	\item Detection rules for vulnerabilities arising from insufficient validation of price data flows. For this category, taint analysis combined with pattern matching is employed. The return values of AMM pools and oracles are defined as taint sources. The corresponding price data flows are then traced to determine whether they undergo the following validation procedures: TWAP smoothing, price-deviation checks, outlier filtering, and multi-source price consistency verification. The abstract syntax tree of the Solidity source code is subsequently examined for the associated vulnerability patterns. The severity of a vulnerability is determined by the number of validation mechanisms detected. The absence of any validation constitutes a high-severity vulnerability, whereas the presence of only one or two validation steps is classified as a medium-severity vulnerability.
	\item Detection rules for oracle invocation vulnerabilities. These vulnerabilities are detected through abstract-syntax-tree pattern matching and access-control analysis. All price-feed update functions declared with external or public visibility are identified, and their access modifiers are examined. The state variables storing oracle addresses are matched, and the functions responsible for updating these addresses are checked for authorization mechanisms such as multi-signature approval or time locks. In addition, the logic for receiving price feeds is analyzed to determine whether timestamp freshness is verified. A price-feed update function without any access restriction, combined with an oracle address that can be arbitrarily modified, is classified as a high-severity vulnerability, whereas the absence of timestamp validation is regarded as a medium-severity vulnerability.
	\item Detection rules for uncontrolled cross-contract invocations. For this category, cross-contract call tracing and return-value analysis are employed to inspect all external calls used to obtain price data. The analysis verifies whether the caller's address is checked against a whitelist before the call and whether the success status and validity of the returned data are verified afterward. The update logic of proxy contracts is also examined to determine whether integrity checks are performed on implementation changes. The absence of caller whitelisting for price-data reception or the lack of return-value validation for external price queries is classified as a high-risk vulnerability, whereas missing integrity checks during proxy implementation upgrades are classified as a medium-risk vulnerability.
	\item Detection rules for AMM price retrieval logic defects. For this category, function-call pattern matching and data-flow analysis are used to identify AMM price retrieval functions within the contract. The analysis determines whether liquidity-depth checks, slippage mitigation for large trades, and aggregation of prices from multiple trading pools are performed before price computation. The absence of liquidity-depth validation, the lack of slippage protection, and the use of a single pool's price as the final price feed are all classified as high-severity vulnerabilities.
\end{enumerate}
To implement the identification method described above, the proposed detection engine is implemented through secondary development based on the open-source static analysis framework Slither. As the most widely used static analysis framework for Solidity smart contracts, Slither provides comprehensive support for abstract syntax tree parsing, control-flow analysis, and data-flow analysis. Its plugin-based architecture enables the rapid implementation of the specialized detection rules proposed in this work, thereby providing a mature technical foundation for the deployment of the detection engine.\\
The engine is implemented as a set of custom Slither plugins and consists of three core modules. The first is a basic information parsing module, which uses Slither's APIs to extract contract metadata and locate key functions. The second is a taint-tracking module, which extends Slither's built-in taint analysis engine with price-related taint-source labeling rules to achieve end-to-end tracking of price data. The third is a rule matching and detection module, which implements the detection rules proposed in this paper to identify vulnerabilities and determine their risk levels. The three modules work in concert to realize the complete detection workflow. The entire analysis can be executed through the \verb+run_detection+ method, which generates a standardized report containing vulnerability details, risk ratings, and remediation recommendations.

\subsubsection{Defense Framework Against Price Manipulation}
\begin{figure}[htbp]
    \centering
    \includegraphics[width=0.8\textwidth]{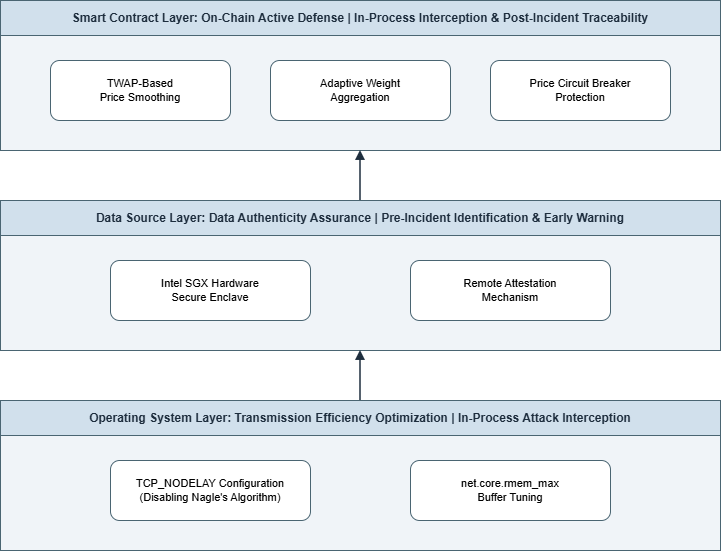}
    \caption{full-stack defense framework}
    \label{fig:full-stack_defense_framework}
\end{figure}

Based on the end-to-end attack paths identified by the attack-tree model and the key risk targets located by the proposed vulnerability identification method, a full-stack layered defense framework is established, encompassing the data source, contract, and operating system layers. By integrating hardware-based trusted execution environments with kernel-level operating system hardening, the framework comprehensively addresses the risks associated with the entire lifecycle of price manipulation attacks.\\
The architecture of the proposed full-stack defense framework is illustrated in Figure~\ref{fig:full-stack_defense_framework}.
At the data source layer, the framework addresses threats such as node compromise at the consensus layer, data-source pollution, and operating-system-level malware attacks. Anchored in the CPU hardware root of trust, a trusted execution environment covering the entire oracle data lifecycle is established using Intel SGX, thereby fundamentally ensuring the authenticity and integrity of price-feed data. Specifically, the \verb+sgx_create_enclave+ instruction is used to create an isolated and encrypted enclave within the CPU. The oracle's core data aggregation algorithms, private keys, and raw data validation procedures are executed inside this enclave. Owing to hardware-level isolation, attempts to tamper with code or data remain ineffective even if the host operating system or kernel is compromised. In addition, remote attestation is employed: before submitting a price feed, each node generates a cryptographic proof of integrity, which is subsequently verified by the on-chain oracle contract. This mechanism guarantees that the data are generated within a trusted environment and remain untampered with throughout the process, thereby preventing malicious node behavior at its source.\\
At the contract layer, to address vulnerabilities at the contract layer and the risk of flash-loan-induced price distortions and short-lived price anomalies within a single block, this layer establishes an active on-chain defense system based on price smoothing, anomaly detection, and automatic circuit breakers.\\
The first component is a multi-source price aggregation and smoothing mechanism. A time-weighted average price (TWAP) is introduced to aggregate prices over multiple consecutive blocks, thereby reducing the impact of transient price fluctuations within a single block and significantly increasing the cost of flash-loan-based manipulation. The second component consists of adaptive weighting and circuit-breaker protection. The influence of oracle nodes is adjusted according to their historical stability, accuracy, and reputation, mitigating the impact of unreliable data sources. In addition, the Uniswap V3 TWAP is adopted as a reference benchmark. If the current price feed deviates excessively from this benchmark, a circuit breaker is triggered to suspend critical contract functions, thereby preventing attackers from profiting and providing administrators with sufficient time to respond.\\
At the operating system layer, the framework focuses on mitigating physical-layer DDoS attacks, price-feed delays caused by network-stack latency, and node synchronization failures. To this end, the Linux kernel network stack is specifically optimized to improve oracle synchronization speed and reduce the attack window.
Two primary measures are adopted. First, the \verb+tcp_nodelay+ kernel parameter is enabled to disable the Nagle algorithm, ensuring that price-feed packets are transmitted without unnecessary delay. Second, the \verb+net.core.rmem_max+ kernel buffer size is appropriately increased to enhance packet reception capacity and reduce packet loss and feed delays caused by kernel processing bottlenecks. These optimizations can reduce network-layer latency to the microsecond level, thereby accelerating synchronization among globally distributed oracle nodes and significantly shortening the time window available for attacks that exploit feed timing discrepancies.

\section{Experiments}
This section conducts empirical evaluations of the price manipulation attack mechanisms, layered defense mechanisms, and vulnerability pattern identification methods proposed earlier. The experiments consist of two major parts. The first part focuses on the reproduction of price manipulation attacks and comparative experiments on defense effectiveness, demonstrating the feasibility of the proposed full-stack layered defense framework. The second part evaluates the smart contract vulnerability pattern identification method, validating its detection capability and advantages. The experiments in this study consist of two parts, each conducted in a different experimental environment to satisfy specific requirements. The defense effectiveness evaluation experiment is implemented using a local single-thread simulation environment based on Python 3.10+, running on Windows 11 and utilizing Matplotlib for visualization. The vulnerability pattern identification experiment is conducted in a static analysis environment built on Ubuntu 22.04 LTS, with related tools including Slither, Solc, and Mythril installed. These two experimental environments provide reliable and consistent platforms for the corresponding experimental tasks.
\subsection{Attack Reproduction and Defense Effectiveness}
\begin{figure}[htbp]
    \centering
    \includegraphics[width=0.8\textwidth]{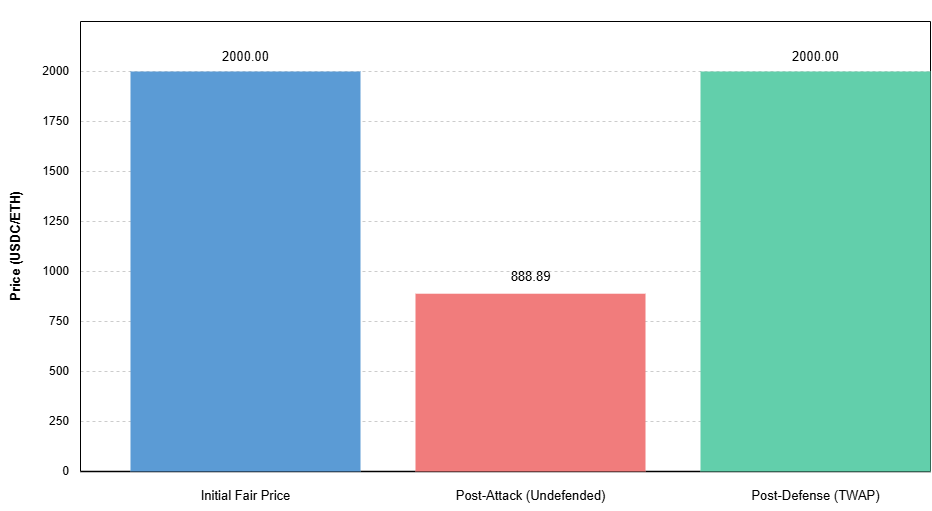}
    \caption{Comparison of AMM Oracle Price Manipulation and Defense Effectiveness}
    \label{fig:quantitative_comparison}
\end{figure}

This study selects a commonly used unprotected native AMM spot oracle as the baseline and compares it with the enhanced defense scheme based on TWAP (Time-Weighted Average Price) and price circuit-breaking mechanisms. Under the same capital scale, AMM pool price manipulation attacks are reproduced, and the effectiveness of the two approaches is compared to verify the proposed defense mechanism.A simplified AMM trading pool based on the constant-product formula is constructed in the local simulation environment. All major parameters are initialized to ensure identical initial conditions for both the control group and the experimental group. The trading pair is simulated using ETH/USDC assets. The initial liquidity is set to 100 ETH and 200,000 USDC, meaning that the initial fair price is 1 ETH = 2,000 USDC. In the enhanced defense scheme, a 10\% price circuit-breaking threshold is configured. Once the absolute deviation between the market price and the TWAP price exceeds this threshold, the oracle price feed is immediately suspended to prevent successful exploitation.\\
The unprotected native oracle used in the control group only provides the capability of retrieving the instantaneous spot price from the AMM pool and incorporates no mechanisms for manipulation prevention or anomaly detection. Under an attack amount of 50 ETH, the reproduced attack causes the oracle price to decline to 888.89 USDC/ETH, resulting in a price deviation of 55.56\%. This demonstrates the inherent risks of native AMM spot-price oracles: attackers only need to provide a proportional amount of capital to significantly alter the pool price, thereby enabling malicious liquidations or the extraction of excessive collateral.\\
The experimental group adopts the enhanced defense scheme proposed in this paper, which establishes a comprehensive protection mechanism based on price smoothing, anomaly detection, and automatic circuit breaking. Under the same attack conditions, although the AMM spot price after the attack decreases to 888.89 USDC/ETH, the TWAP price after applying the defense mechanism remains close to the original fair price. The price deviation is maintained below 5\%, and the circuit breaker is successfully triggered. This demonstrates that the proposed defense scheme can effectively mitigate price manipulation attacks and prevent attackers from exploiting the vulnerability in a timely manner.\\
The quantitative comparison between attack reproduction and defense effectiveness is illustrated in Figure~\ref{fig:quantitative_comparison}
To address the additional computational overhead introduced by the defense mechanisms, comprehensive testing was conducted throughout this study. The results show that the TWAP mechanism and the circuit-breaker mechanism introduce only a marginal increase in operations, which has little impact on the oracle's price-feeding performance under normal conditions. When combined with kernel-level network optimization, the attack window can be further narrowed, achieving a favorable trade-off between security and efficiency.

\subsection{Vulnerability Pattern Recognition}
\begin{figure}[htbp]
    \centering
    \includegraphics[width=0.8\textwidth]{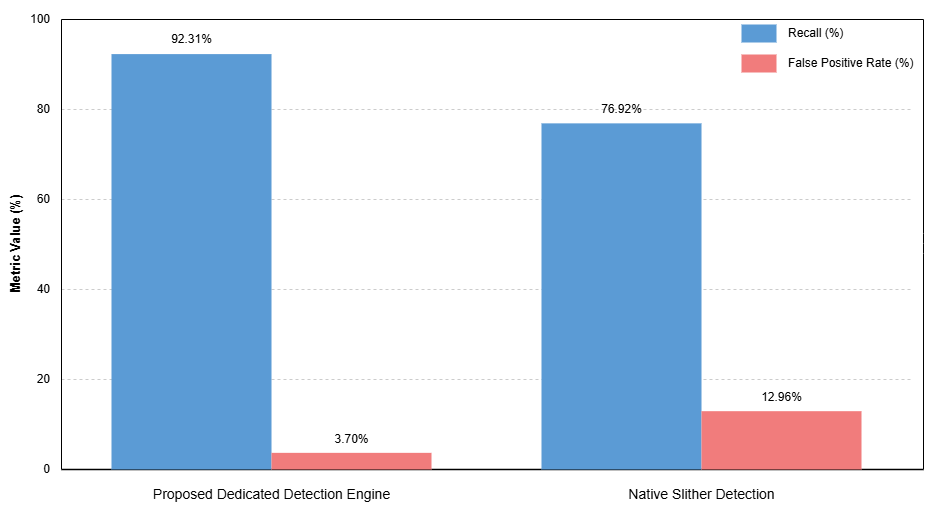}
    \caption{Performance Comparison of Different Detection Tools}
    \label{fig:Performance_Comparison}
\end{figure}

Drawing on Ethereum blockchain explorers, open-source repositories of DeFi attack incidents, and existing smart contract vulnerability datasets, this study collected 52 real-world contract instances exhibiting price-manipulation-related vulnerabilities, together with 108 security-audited contracts free of such issues serving as a control group, yielding a standardized dataset of 160 contracts in total. This dataset encompasses all four categories of price manipulation vulnerability patterns discussed above, as well as DeFi contracts written in different Solidity versions and spanning different business types, thereby enhancing the representativeness and fairness of the experiments.\\
The static analysis environment for this experiment was built on Ubuntu 22.04 LTS, in which the dedicated detection engine proposed in this study was deployed alongside two widely used industry detection tools, Slither and Mythril, for horizontal comparison. Detection effectiveness and performance were evaluated primarily using four metrics: precision, recall, false positive rate, and detection time. Precision reflects the overall accuracy of the detection process, recall reflects the proportion of actual vulnerabilities successfully identified, the false positive rate reflects the proportion of contracts erroneously flagged as insecure, and detection time reflects the execution speed of the method.\\
Detection experiments were conducted on the test dataset. For price-manipulation-related vulnerability detection, the dedicated detection engine designed in this study achieved a precision of 94.38\% and a recall of 92.31\%, with an average detection time of 128 ms per contract. Notably, the false positive rate for targeted detection of high-risk vulnerabilities was as low as 3.70\%, while the overall false positive rate across all vulnerability types was 4.63\%. These experimental results demonstrate that the recognition method proposed in this study can accurately and efficiently identify price-manipulation-related vulnerability patterns in smart contracts, exhibiting excellent capability in identifying high-risk vulnerabilities while maintaining an extremely low false positive rate, thereby demonstrating strong practical application value.\\
Under identical experimental conditions and using the same dataset, the dedicated detection engine designed in this study was compared with mainstream industry tools in terms of core performance metrics. The results are shown in Figure~\ref{fig:Performance_Comparison}.\\

\section{Conclusion}
This paper conducts a comprehensive study of AMM spot-oracle price manipulation attacks within the DeFi ecosystem. It first introduces the relevant background concepts and technologies, including the basic architecture of blockchain, the AMM constant-product model, oracles, and flash loans, thereby clarifying the underlying technical enablers of such attacks and the security threats they pose across the entire oracle lifecycle. In this study, we build on that foundation by proposing a three-layer attack tree model to explain the attack mechanism. The model consists of three layers: the physical/infrastructure layer, the protocol/consensus layer, and the application/logic layer. Through this framework, we identify the oracle's "data election" stage as the optimal point for an attacker to inject an attack. Four categories of smart contract vulnerability are then defined: missing price data flow validation, risky oracle invocation, uncontrolled cross-contract invocation, and defective AMM price-reading logic, each illustrated through concrete case studies. On this basis, a fuzzy analytic hierarchy process (FAHP) is employed to construct a comprehensive vulnerability risk assessment model spanning the technical, market, governance, and contract-code dimensions. An automated vulnerability detection method based on Slither is further proposed, together with a multi-layered defense scheme covering the data-source layer, the contract layer, and the operating-system layer. Simulation results show that the proposed defense measures can effectively limit the post-attack price deviation to no more than 5\%, while the proposed detection method achieves high precision and recall, meeting the anticipated objectives and demonstrating practical significance for the security protection of DeFi oracles.\\

\section{Acknowledgments}
AI-based tools are used for language polishing during manuscript preparation. 
\bibliographystyle{unsrtnat}
\bibliography{references}

\end{document}